\documentclass[sigconf,natbib=true,anonymous=false]{acmart}

\usepackage{amsmath,nccmath,tabularx} 
\usepackage{algorithm}
\usepackage{algorithmic}
\usepackage{color}
\usepackage{graphicx}
\graphicspath{{images/}}
\usepackage{wrapfig,lipsum}
\usepackage{tabularx}
\usepackage{graphicx}
\usepackage{lscape}
\usepackage{subcaption}
\usepackage{latexsym}
\usepackage{balance}
\usepackage{pifont}
\usepackage{multirow}

\usepackage{enumitem}
\setlist{leftmargin=2mm}
\usepackage[T1]{fontenc}
\usepackage[utf8]{inputenc}
\usepackage[font=small,labelfont=bf]{caption}

\AtBeginDocument{%
  \providecommand\BibTeX{{%
    \normalfont B\kern-0.5em{\scshape i\kern-0.25em b}\kern-0.8em\TeX}}}

\copyrightyear{2024}
\acmYear{2024}
\setcopyright{licensedothergov}\acmConference[SIGIR '24]{Proceedings of the 47th International ACM SIGIR Conference on Research and Development in Information Retrieval}{July 14--18, 2024}{Washington, DC, USA}
\acmBooktitle{Proceedings of the 47th International ACM SIGIR Conference on Research and Development in Information Retrieval (SIGIR '24), July 14--18, 2024, Washington, DC, USA}
\acmDOI{10.1145/3626772.3657798}
\acmISBN{979-8-4007-0431-4/24/07}
\acmPrice{}
\begin{document}

\title{Leveraging LLMs for Unsupervised Dense Retriever Ranking}
\author{Ekaterina Khramtsova}
\authornote{Equal Contribution}
\affiliation{%
	\institution{University of Queensland}
	\streetaddress{4072 St Lucia}
	\city{St Lucia}
	\country{Australia}}
\email{e.khramtsova@uq.edu.au}

\author{Shengyao Zhuang}
\authornotemark[1]
\affiliation{%
	\institution{CSIRO}
	\streetaddress{4072 St Lucia}
	\city{Herston}
	\country{Australia}}
\email{shengyao.zhuang@csiro.au}

\author{Mahsa~Baktashmotlagh}
\affiliation{%
	\institution{University of Queensland}
	\streetaddress{4072 St Lucia}
	\city{St Lucia}
	\country{Australia}}
\email{m.baktashmotlagh@uq.edu.au}

\author{Guido Zuccon}
\affiliation{%
	\institution{University of Queensland}
	\streetaddress{4072 St Lucia}
	\city{St Lucia}
	\country{Australia}}
\email{g.zuccon@uq.edu.au}






\begin{abstract}

 In this paper we present Large Language Model Assisted Retrieval Model Ranking (LARMOR), an effective unsupervised approach that leverages LLMs for selecting which dense retriever to use on a test corpus (target). Dense retriever selection is crucial for many IR applications that rely on using dense retrievers trained on public corpora to encode or search a new, private target corpus. This is because when confronted with domain shift, where the downstream corpora, domains, or tasks of the target corpus differ from the domain/task the dense retriever was trained on, its performance often drops. Furthermore, when the target corpus is unlabeled, e.g., in a zero-shot scenario, the direct evaluation of the model on the target corpus becomes unfeasible. Unsupervised selection of the most effective pre-trained dense retriever becomes then a crucial challenge. Current methods for dense retriever selection are insufficient in handling scenarios with domain shift.

Our proposed solution leverages LLMs to generate pseudo-relevant queries, labels and reference lists based on a set of documents sampled from the target corpus. Dense retrievers are then ranked based on their effectiveness on these generated pseudo-relevant signals. Notably, our method is the first approach that relies solely on the target corpus, eliminating the need for both training corpora and test labels. To evaluate the effectiveness of our method, we construct a large pool of state-of-the-art dense retrievers. The proposed approach outperforms existing baselines with respect to both dense retriever selection and ranking. We make our code and results publicly available at \url{https://github.com/ielab/larmor/}.

\end{abstract}

\vspace{-10pt}

\vspace{-4pt}
\keywords{Model selection, Dense retrievers, Zero Shot Model Evaluation \vspace{-4pt}}

\maketitle

\section{Introduction}

\begin{figure}
    \centering
    \includegraphics[width=0.45\textwidth]{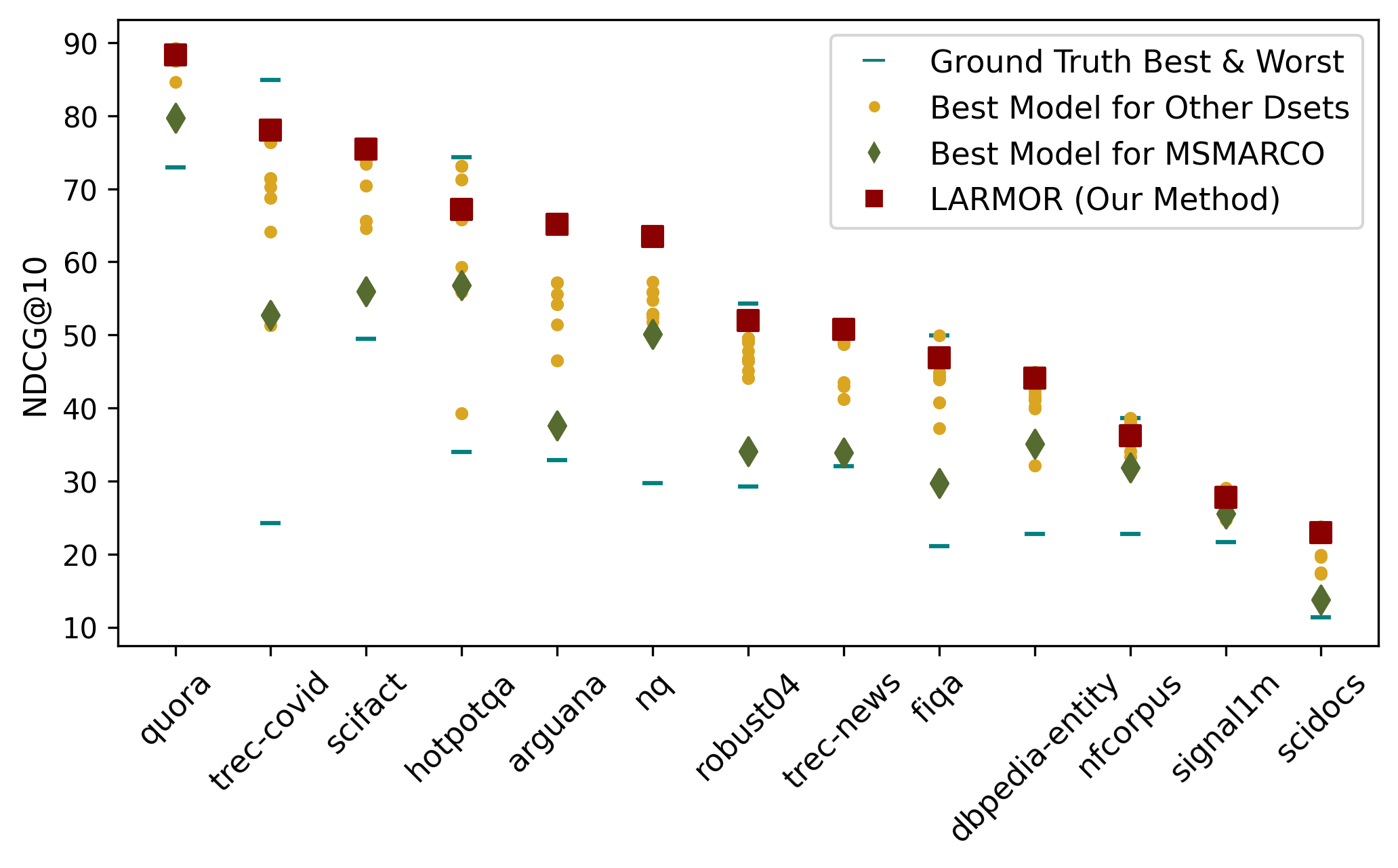}
    \vspace{-12pt}
    \caption{nDCG@10 Across Collections. This figure illustrates that selecting the best model based on one collection (as indicated by orange dots) does not necessarily 
    ensure its effectiveness on another.
    In contrast, our unsupervised approach (indicated by red squares) consistently selects competitive models across various collections, and even identifies the most performant model for Trec-News.\vspace{-6pt}}
    \label{ndcg_scatter}
    \vspace{-10pt}
 \end{figure}

With the rapid advancements in natural language processing and information retrieval, a multitude of diverse dense retrievers (DRs) has emerged, demonstrating impressive effectiveness in various text retrieval tasks~\cite{tonellotto2022lecture,10.1145/3637870}. For example, over 100 dense retrievers are featured in the Massive Text Embedding Benchmark\footnote{\url{https://huggingface.co/spaces/mteb/leaderboard}} (MTEB~\cite{muennighoff-etal-2023-mteb}).

A typical situation in practical application settings is that a dense retriever trained on one or more corpora (the \textit{training corpus}) is to be applied to a new corpus (the \textit{target corpus}). Often queries and relevance judgements (labels) for the target corpus are not available, or they are prohibitive to collect due to costs or data access restrictions (this is often the case in domain specific settings like in health, legal and patent search). In this situation, then, search engine practitioners are faced with the question --- Which dense retriever should I use? This is the task of \textit{dense retriever selection}~\cite{khramtsova2023}:  identify the most suitable DR for a target corpus. This is a challenging task because no queries and associated relevance judgements are available for the target corpus, and thus the prediction task is to be performed in an unsupervised manner.

\begin{table*}[t]
    \caption{nDCG@10 on BEIR. 47 top-performing DRs from METB are used in this experiment. The first row (Oracle) reports the scores obtained when using the best DR for each collection, representing the upper bound score. The second row (Best DR) reports the scores of \texttt{UAE}, the DR that achieved the best average nDCG@10 on BEIR. The last row (LARMOR) reports the scores achieved using DRs selected by our  method. \vspace{-6pt}}
\resizebox{\textwidth}{!}{
\centering
\begin{tabular}{c|c|c|c|c|c|c|c|c|c|c|c|c|c|c}
    \toprule
    &  NF & FiQA & ArguAna & SciDocs & SciFact& Covid & Quora&  NQ & DBPedia & HotpotQA  & Signal1M & Robust04 & Trec-News &\textbf{Avrg}\\
    \midrule

    Oracle (Upper Bound) & 38.65 & 49.96 & 65.11 & 23.77 & 76.18 & 84.88 & 89.26 & 64.07 & 44.89 & 74.33 & 29.04 & 54.31 & 50.77 & 57.32\\
    Best DR (\texttt{UAE})&   38.65 & 44.84 & 65.11 & 22.98 & 74.07 & 76.33 & 88.79 & 55.86 & 44.89 & 73.13 & 27.36 & 49.55 & 49.21 & 54.67 \\
    LARMOR (ours)& 36.21 & 46.89 & 65.11 & 22.98 & 75.41 & 78.07 & 88.32 & 63.49 & 44.07 & 67.16 & 27.76 & 51.94 & 50.77& 55.24\\
    \bottomrule
\end{tabular}
}
\vspace{-7pt}
\label{tab:gt_ndcg_vs_us}
\end{table*}

A reasonable choice for dense retriever selection would be to select the DR that performs overall best on a comprehensive leaderboard like MTEB.
However, recent studies have shown that the effectiveness of DRs is often dependent on the similarity between the training corpus and the target corpus; in particular, the effectiveness becomes varying and unpredictable when DRs are applied to data that differs from that at training (e.g., from a new domain, see Figure~\ref*{ndcg_scatter})~\cite{thakur2021,zhuang2021dealing,zhuang2022char,ren-etal-2023-thorough,lin2023train,Lupart2022MSShiftAA}. 
This issue is evident for instance in the MTEB benchmark. There, results show that the overall top-performing DRs may not necessarily be the most suitable for each single collection.
For instance, in our experiments with a subset of the DRs from the MTEB benchmark, \texttt{all-mpnet-base-v2}\footnote{\url{https://huggingface.co/sentence-transformers/all-mpnet-base-v2}} is the top performing dense retriever on FiQA (nDCG@10 = 0.4996). 
In contrast, the overall leading dense retriever on the MTEB benchmark is \texttt{UAE-Large-V1}\footnote{\url{https://huggingface.co/WhereIsAI/UAE-Large-V1}}, which on FiQA exhibits a significant 10.3\% loss in nDCG@10 compared to \texttt{all-mpnet-base-v2}.

Another straightforward approach to DR selection would be to select the DR that performs best on a held-out portion of the training corpus. This, has been shown to be the most effective method for selecting DRs in previous work~\cite{khramtsova2023}. However, a significant difficulty arises when doing this: new state-of-the-art DRs are often trained on multiple, proprietary, corpora, e.g., \texttt{e5}~\cite{wang2022text}. This renders access to training and/or held-out data impractical or impossible.

Other alternatives have been recently explored, adapted from similar problems in computer vision ~\cite{khramtsova2023}. These, however, necessitate the availability of queries from the target corpus. This requirement poses a practical challenge in real-world scenarios, where the decision on which DR model to use must be made prior to deploying the application and thus often no prior logged queries are available. Nevertheless, even if logged queries are available, these approaches have been shown largely ineffective for DR selection ~\cite{khramtsova2023}.

In this paper, we propose a family of approaches for unsupervised, query-free dense retriever selection. At the core of these approaches is the leveraging of the capabilities of Large Language Models (LLMs)~\cite{zhao2023survey}. 
Specifically, we address the challenge posed by the absence of queries by using LLMs to generate synthetic queries for a (subset of the) target corpus. A document for which a synthetic query is generated and the generated query itself are considered forming a pseudo-relevant query-document pair for the target corpus. The set of  pseudo-relevant query-document pairs are then used to estimate the ranking effectiveness of the DRs on the target corpus, and in turn this is used to rank DR systems. 

Our results demonstrate that this straightforward performance estimation based on query generation  is remarkably effective in selecting the most suitable DR for a target corpus -- outperforming any other DR selection method.
We further propose refinements to this idea that encompass the generation of synthetic relevance labels, and the exploitation of synthetic reference lists. The combination of these methods leads to a highly effective unsupervised strategy for dense retriever selection, which we refer to as Large Language Model Assisted Retrieval Model Ranking (LARMOR). 

Table~\ref{tab:gt_ndcg_vs_us} provides a snapshot of LARMOR's 
predictive capabilities when selecting DRs for a target corpus: this serves as a motivation to delve further into the remainder of the paper. Each column in the table represents a target corpus
(the last column is the mean effectiveness), and the value reported is the effectiveness on the target corpus of the selected dense retriever (fine-tuned on a different training corpus); Section~\ref{sec:setup} details our empirical settings. The first row (Oracle) refers to the best performance attainable if the effectiveness of every dense retriever on each of the target corpora were known -- this is a theoretical upper bound. The second row reports the performance attainable when selecting the single model that performs overall best across all considered target corpora; in the case of the table, such a model is \texttt{UAE}. Again, this method is impossible in practice as it requires the true relevance labels for each target corpus to determine the overall best DR. Finally, the third row reports the remarkable performance of our LARMOR: these were obtained without resorting to human annotations, nor access to queries from the target corpus, which are often unfeasible to obtain before deployment. LARMOR in fact manages to select a highly competitive DR for each of the target corpora, and overall LARMOR provides better DR selection than using the  \texttt{UAE} model across all target corpora -- recall than \texttt{UAE} could have been selected only because we accessed the relevance labels of each target corpus. 
In addition, LARMOR performance is only 3.6\% less than the theoretical upper bound (Oracle).

While 
our primary focus is on the DRs, our method can be applied to other IR models (e.g. re-rankers or sparse models).
LARMOR is now integrated in the \texttt{DenseQuest}~\cite{khramtsova2024densequest} system that implements DR selection over custom collections (\url{https://densequest.ielab.io})

\noindent\textbf{Key contributions:}


\begin{enumerate}

\vspace{-0.7ex}
\item We introduce Large Language Model Assisted Retrieval Model Ranking (LARMOR), an approach for dense retriever selection that exploits the zero-shot capability of LLMs for generation of queries, relevance judgments, and reference lists. LARMOR is highly effective in selecting a dense retriever for a target corpus, without the need to supply queries or labels from the target corpus.

\item To assess LARMOR's performance, we assemble a pool of 47 top DRs from the MTEB retrieval benchmark, extending the results of previous work to considering a consistently larger set of models.

\item We conduct a thorough ablation study to examine factors that impact LARMOR's effectiveness, including the type and size of LLMs used, and the number of generated queries per documents.

\item We augment the set of baselines for dense retriever selection by evaluating existing query performance prediction (QPP) methods overlooked by previous work~\cite{khramtsova2023}.
\end{enumerate}



\section{Related work}

\subsection{Dense Retrievers Selection}

While the task of unsupervised model selection has been widely studied for general deep learning models ~\cite{Guillory2021PredictingWC,ATC, deng2020labels, chenmandoline2021, pmlr-v206-khramtsova23a}, its application in the context of neural rankers remains largely unexplored. The recent study by Khramtsova et al.~\cite{khramtsova2023} formalized the problem of DR selection and proposed several baseline methods. However, their results indicated that most methods adapted from other areas are ineffective for IR. Differing from ~\cite{khramtsova2023}, we introduce a more challenging experimental setup by expanding the number of DRs and adding an additional constraint: the models might have different training sets. Consequently, we had to exclude several baselines, namely query similarity and Fr\'echet-based corpus similarity, as they require access to the training data. We retain the other baselines for our comparison—\texttt{MSMARCO perf}, \texttt{Binary entropy}, and \texttt{Query Alteration}—using the best-reported hyperparameters.

\vspace{-5pt}
\subsection{Query Performance Prediction}


The concept of performance estimation in IR is primarily investigated within the context of query performance prediction (QPP)~\cite{carmel2012query, hauff2008survey, he2006query}. QPP aims to predict the retrieval quality of a search system for each query independently, without relying on relevance judgments. Our paper's objective differs slightly, focusing on comparing performance across different rankers, rather than evaluating each query within a single ranker. Nonetheless, it seems logical to explore the adaptation of QPPs to the task of DR selection.

Traditional QPP methods fall into three main categories~\cite{Shtok2016QueryPP}: those that assess the clarity of search results relative to the overall corpus; those that analyze the retrieval scores of documents within the ranking lists; and those that evaluate the robustness of the predicted ranking.
We adapt methods from each of these categories to the DR selection task. From the first category, we employ \texttt{Clarity}~\cite{Cronen2002Clarity}; from the second, we utilize \texttt{WIG}~\cite{Yun2007Wig}, \texttt{NQC}~\cite{Shtok2012NQC}, \texttt{SMV}~\cite{Tao2014SMV}, and \texttt{$\sigma$}~\cite{Perez2010Sigma,Cummins2011SigmaMax}; and from the third category, we explore \texttt{Fusion}~\cite{Shtok2016QueryPP}. 
Additionally, \texttt{Query Alteration}~\cite{khramtsova2023} can also be considered as an adaptation of robustness-based QPP. Our LARMOR method bears similarities with the concurrently proposed QPP-GenRE~\cite{meng2024query}; however QPP-GenRE is intended for the QPP task and it only considers producing synthetic labels, not queries.

\vspace{-4pt}
\subsection{Challenges For The Existing Baselines} \label{sec:baseline_challenges}

Next we discuss the challenges encountered in adapting existing methods to our task of unsupervised DR ranking and selection.

\subsubsection{Normalizing factors for score-based methods.}
    Score-based QPP methods 
 can be significantly enhanced by scaling their respective query-based measures with the relevance of the entire corpus to that query~\cite{Kurland2012QPP, Shtok2016QueryPP}. In the context of DRs, this scaling is akin to the score between the query and the entire target corpus; however, calculating this value is computationally unfeasible~\cite{Faggioli2023QPP}. 
     Several methods have been proposed to approximate this scaling parameter. One approach, as suggested by Meng et.al.~\cite{Meng2023ConvSearchQPP}, is to take the average score of the top retrieved documents for each query as the normalizing factor. Another method, defined by Faggioli et.al.~\cite{Faggioli2023QPP}, involves representing the entire corpus as a centroid of its documents, derived from the latent space of the DR. 
     The normalizing factor is then calculated as the score between the query and this centroid representation. We followed \cite{Meng2023ConvSearchQPP} in our experiments, and report both variations with and without normalization.

     Typically, the scores generated for a query-document pair do not accurately reflect the probability of the query being relevant to the document. For example, in models trained using the dot product, the scores are not bound to a range between 0 and 1.

     In addition to the variability of the scores  within one collection, 
     another challenge is the variability of the score distributions across DRs.
     This variability arises due to each dense retriever having unique architecture, loss function, and other training hyperparameters. As a result, even by normalizing the scores within one corpus, score-based methods do not perform well when used for comparing different dense retrievers, as will be shown in the next section.

\subsubsection{Variability of pre-trained tasks and training collections for performance-based methods.}
DRs are seldom trained from scratch; rather, it is common to start with a model pre-trained on a different task and fine-tune it for a retrieval task. Consequently, it is logical to use the performance on the original task and the new retrieval task as indicators of model generalizability. For instance, if a model was pre-trained on a masked language modeling task (such as BERT~\cite{devlin2018bert} and Roberta~\cite{liu2019roberta}),
    one could evaluate the model's adaptability to a new task by examining its robustness to masking, as demonstrated in the Query Alteration Method. Similarly, if a model was fine-tuned on MSMARCO, its performance on MSMARCO can serve as an indicator of its ability to perform a retrieval task.

    However, a challenge emerges because the dense retrievers in our study are not pre-trained using the same task. For instance, while some models are based on BERT, others are built on GPT~\cite{brown2020language} 
     and were initially pre-trained using Next Token Prediction. Additionally, the training corpora differ across models. This diversity results in a performance evaluation that may be biased towards models trained on specific tasks or corpora, potentially not reflecting true performance on the target corpus or task.

\begin{figure*}
	\centering
	\includegraphics[width=0.99\textwidth]{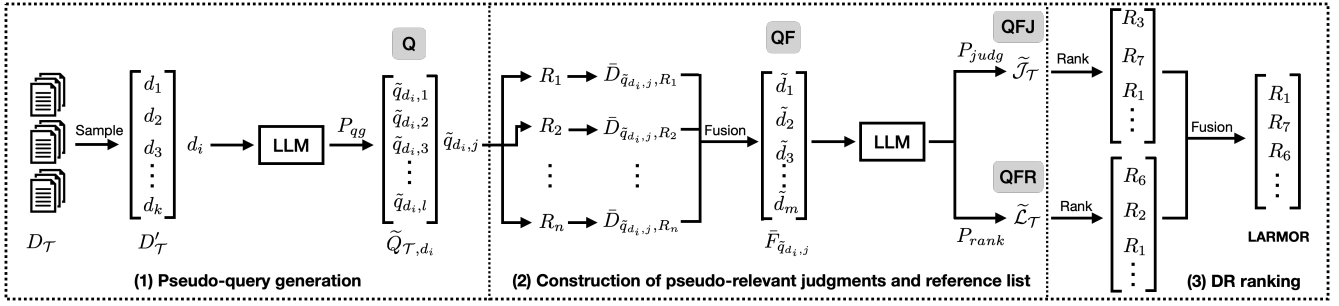}
	\vspace{-10pt}
	\caption{The LARMOR dense retriever selection pipeline. Labels Q, QF, QFJ and QFR refer to the ablation points described in Section~\ref{sec:ablation}.}
	\label{pipeline}
	\vspace{-8pt}
\end{figure*}

\vspace{-5pt}
\subsection{LLMs for Information Retrieval}

In our research, we rely on LLMs to generate pseudo-relevant queries, labels, and reference lists in the zero-shot setting. Numerous papers in the NLP and IR literature  demonstrated the remarkable zero-shot performance of LLMs in these tasks.

For query generation, previous methods focused on fine-tuning pre-trained language models to generate pseudo-relevant queries~\cite{nogueira2019doc2query, gospodinov2023doc2query}. More recently, many works demonstrated that it is possible to generate high-quality queries for training ranking models~\cite{inpars, inparsv2, dai2022promptagator, wang2021gpl} by prompting LLMs in a zero-shot manner. Hence, following these works, we also use a domain-specific query generation prompt to guide LLMs in generating pseudo-relevant queries. 

Another important component in our pipelines is the generation of pseudo-relevant judgments for the generated queries. \citeauthor{thomas2023large}~\cite{thomas2023large} demonstrate that using GPT-4 to generate relevant judgments can surpass human annotators in evaluating search systems. Similarly, a study conducted by \citeauthor{guglielmo2023perspectives}~\cite{guglielmo2023perspectives} shows that LLM-generated relevant labels can align with labels annotated by TREC annotators. Thus, in this work, we leverage LLMs to provide additional relevance judgments in addition to the generated queries.

Finally, our work also leverages a pseudo-reference list to evaluate the effectiveness of DRs. This involves generating document rankings with high ranking scores (e.g., high nDCG). In IR, numerous works have demonstrated that LLMs have very strong zero-shot ranking capabilities. The methodologies for harnessing LLMs in zero-shot ranking tasks can be broadly categorized into pointwise~\cite{zhuang-etal-2023-open, sachan-etal-2022-improving, zhuang2023beyond}, listwise~\cite{sun-etal-2023-chatgpt, ma2023zero, pradeep2023rankvicuna, pradeep2023rankzephyr, tamber2023scaling}, pairwise~\cite{qin2023large}, and setwise~\cite{zhuang2023setwise}. In our paper, we adopt a setwise approach to generate a pseudo-reference list due to its high effectiveness and efficiency.

\vspace{-4pt}
\section{Problem formulation}

Let $\mathcal{T}$ be a target collection containing a corpus $D_\mathcal{T}$ of documents, a set $Q_\mathcal{T}$ of queries, and a set $\mathcal{J}_{Q_\mathcal{T},D_\mathcal{T}}$ of relevance judgments (i.e., labels), which reflect the degree of relevance of a given document  $d \in D_\mathcal{T}$ in relation to a specific query $q \in Q_\mathcal{T}$.

We note that accessing such relevance judgments presents significant challenges: queries often contain private user information; while collecting high-quality relevance judgments is time-consuming, and for some applications (e.g., medical or legal IR) requires in-depth domain knowledge and thus can be costly. 

Let $\mathcal{R} = \{R_1, R_2, \ldots R_n\}$ be a set of rankers, each trained on its respective training collection\footnote{Multiple training collections could also be used to derive a dense retriever: this does not change the setup discussed here.
}: $\mathcal{S} = \{\mathcal{S}_{R_1},\mathcal{S}_{R_2}, \ldots ,\mathcal{S}_{R_n} \}$.  
Note that the target collection $\mathcal{T}$ was not used during training of $\mathcal{R}$: $\mathcal{T} \nsubseteq \mathcal{S}$. 
In current applications, the training collections $\mathcal{S}_{R_i}$ often contain large corpora, which often include private documents -- e.g., those used to train the latest state-of-the-art dense retrievers like \texttt{e5}~\cite{wang2022text}. Therefore, we operate under the assumption that only the trained dense retrievers in $\mathcal{R}$ are available, while access to $\mathcal{S}$ is restricted.

Finally, let $\mathcal{E}$ be an evaluation measure, such as nDCG@10. Practitioners can establish an ordering of DRs in $\mathcal{R}$ based on the value of the evaluation measure $\mathcal{E}$ obtained on the target collection. This is achieved by applying each DR to the target collection and using the relevance judgments to compute the evaluation measure. The rankers are then arranged in decreasing order of $\mathcal{E}$, creating the ranking (ordering) of DRs $\mathcal{O(R, T, E, J)}$. The top-ranked DR is then typically selected for deployment as a search function on the target corpus of documents $D_\mathcal{T}$ to answer new queries, as it is the one that has been found performing best on the target collection $\mathcal{T}$. We note that we consider $\mathcal{O(R, T, E, J)}$ to be the ground truth ranking for the dense retriever selection task, defined below, since the DRs are evaluated and ranked based on the ground truth relevance judgments.

\textit{Dense retriever selection task}: The problem of dense retriever selection consists of predicting the ranking $\mathcal{O(R, T, E, J)}$ without accessing the relevance judgments $\mathcal{J}_{Q_\mathcal{T},D_\mathcal{T}}$, as well as the target queries $Q_\mathcal{T}$. This is equivalent to producing a ranking $\hat{O}(\mathcal{R}, D_\mathcal{T}, \mathcal{E})$ of the rankers in $\mathcal{R}$ for the target collection $\mathcal{T}$ and evaluation measure $\mathcal{E}$, such that $\hat{O}(\mathcal{R}, D_\mathcal{T}, \mathcal{E})$ corresponds to the true ranking $\mathcal{O(R, T, E, J)}$. Note that $\hat{O}(\mathcal{R}, D_\mathcal{T}, \mathcal{E})$ does not include the relevance assessments $\mathcal{J}_{Q_\mathcal{T},D_\mathcal{T}}$ and target queries $Q_\mathcal{T}$ as input. Our goal is to develop a dense retriever selection method $M(\mathcal{R}, D_\mathcal{T})$ that produces the ranking $\hat{O}(\mathcal{R}, D_\mathcal{T}, \mathcal{E})$.

\vspace{-6pt}
\section{Methodology}
Next we describe our method,  Large Language Model Assisted Retrieval Model Ranking (LARMOR). LARMOR uses Large Language Models (LLMs) to tackle the dense retriever selection problem. The method encapsulates a pipeline consisting on three crucial components, as outlined in Figure~\ref{pipeline}: (1) pseudo-query generation, (2) construction of pseudo-relevant judgments and reference lists, and (3) dense retrievers ranking.

\vspace{-4pt}
\subsection{Pseudo-Query Generation}
The first step of the LARMOR pipeline involves generating pseudo-queries for the target corpus $D_\mathcal{T}$ to address the challenge of the absence (i.e., inability to access) of a representative set of queries for the target collection, $Q_\mathcal{T}$.

For this, we start by randomly sampling a subset of $k$ documents from the target corpus:  $D'_\mathcal{T} =\{d_1, \ldotp, d_k\} \in D_\mathcal{T}$. 
Once the subset $D'_\mathcal{T}$ is built, each document from $D'_\mathcal{T}$ is passed to LARMOR's LLM, accompanied by a domain-specific query generation prompt $P_{qg}$, to generate a set of pseudo-relevant queries $\widetilde{Q}_\mathcal{T}$ specific to the target collection\footnote{That is, we assume that it is known what the representative search tasks for the target collection are.}. Note that for each sampled document $d_i$, we generate $l$ queries ($\tilde{q}_{d_i,1}, \dots \tilde{q}_{d_i,l}$):
\vspace{-6pt}
\begin{equation}\small
	\widetilde{Q}_\mathcal{T} =  \bigcup_{i=1}^{k} \prod_{j=1}^{l} LLM(P_{qg}(d_i)).
	\vspace{-4pt}
\end{equation}
\noindent in the equation, $\prod_{l}$ symbolises the generation of $l$ queries for a single document $d_i$, each time using the same prompt template $P_{qg}$ specific to the domain of the target collection. 

For instance, in the case of the target domain being Wikipedia (e.g., for the NQ corpus in BEIR), the query generation prompt we employ is \textit{``Generate a question that the following Wikipedia page can answer. Avoid generating general questions. Wikipedia page: $\{d_i\}$''}, where ${ d_i}$ is a placeholder for the sample document text. By inputting this prompt to the LLM, it is possible to generate in-domain queries for the collection, thus addressing the challenge of the unavailability of a target query set $Q_\mathcal{T}$. This prompt design requires only minimal prior knowledge about the collection domain, effectively mimicking real-world scenarios.
The design of prompts specific to a target collection has been shown effective in previous work in the context of training dense retrievers~\cite{asai-etal-2023-task,zhang2023retrieve,wang2022text}. To the best of our knowledge, we are the first to design and use domain-specific prompts for query generation.

It is worth noting that generating multiple pseudo-relevant queries for each sample document\footnote{Recall that we generate $l$ queries for each sample document.} using sampling generation strategies, such as top-p sampling~\cite{holtzman2019curious}, is reasonable. Since one document is likely to cover different topics, it could be relevant to various queries. Generating multiple queries with sampling strategies has the potential to cover different aspects of the document.

\subsection{Construction of Pseudo Relevance Judgments and Reference Lists}
\label{sec:psuedo-jud-list}
Once $\widetilde{Q}_\mathcal{T}$ is obtained, we could construct pseudo-relevant judgments to evaluate dense retrievers in the candidate pool by assuming a document is relevant to its corresponding generated queries.
However, such an approach can only provide one relevant document per generated query, and these shallow relevance judgments may be sub-optimal for evaluating the ranking effectiveness of a DR~\cite{mackenzie2021sensitivity,10.1007/s10791-022-09411-0,10.1145/3239572,lu2016effect}. Consequently, the next step in the LARMOR pipeline is to construct comprehensive pseudo-relevant signals to evaluate and rank DRs in the candidate pool. We propose two different types of pseudo-relevant signals for this purpose: pseudo-relevant judgments and pseudo-reference lists, each with a distinct way of prompting LLMs and evaluating DRs.

For both pseudo-relevant signals types we start by creating a single document ranking for each generated query $\tilde{q}_{d_i,j}  \in \widetilde{Q}_{\mathcal{T},d_i}$, where $\widetilde{Q}_{\mathcal{T},d_i}$ is the subset of $\widetilde{Q}_{\mathcal{T}}$ that contains the $l$ generated queries for sampled document $d_i$. 
The single document ranking for the generate query $\tilde{q}_{d_i,j}$ is achieved by submitting the query to all considered DRs in $\mathcal{R} = \{R_1, R_2, ..., R_n\}$, to obtain the document rankings $\{\widebar{D}_{\tilde{q}_{d_i,j},R_1}, \widebar{D}_{\tilde{q}_{d_i,j},R_2}, ..., \widebar{D}_{\tilde{q}_{d_i,j},R_n}\}$. Here, $\widebar{D}_{\tilde{q}_{d_i,j},R_1}$ is the ranking of documents in the corpus $D_{\mathcal{T}}$ induced by dense retriever $R_1$ for query $\tilde{q}_{d_i,j}$. While the notation assumes that any $\widebar{D}_{\tilde{q}_{d_i,j},R}$ is a total ordering of the documents in $D_{\mathcal{T}}$, in practice retrieval is conducted up to a rank cut-off (typically 1,000): the cut-off value has little effect on the result of the fusion, provided it is large enough (i.e. larger than parameter $m$ below, the number of documents selected from the fused ranking).
Subsequently, a rank fusion algorithm\footnote{Which rank fusion method to use is an implementation choice; many  exist~\cite{kurland2018fusion}.} is employed to merge all the rankings for $\tilde{q}_{d_i,j}$, resulting in a single fused document ranking from which we select only the top-$m$ documents, obtaining the document ranking $\widebar{F}_{\tilde{q}_{d_i,j}}$ of size $m$. One can consider this step as selecting the most valuable $m$ documents for query $\tilde{q}_{d_i,j}$ for the LLM to judge or rank: the documents in $\widebar{F}_{\tilde{q}_{d_i,j}}$ are likely retrieved by most of the DRs. Hence, providing relevance judgments or reference lists for these documents may yield a more accurate evaluation of the DR effectiveness.

Next, $\widebar{F}_{\tilde{q}_{d_i,j}}$ is passed as input to the LLM using either (or both) of two prompts (if both, this is done separately, i.e. independent inferences for each prompt). 

Prompt $P_{judg}$ instructs the LLM to generate the pseudo-relevance judgments\footnote{Sometimes referred to as synthetic judgements.} 
$\widetilde{\mathcal{J}}_{\tilde{q}_{d_i,j}, \widebar{F}_{\tilde{q}_{d_i,j}}}$ for the m documents in $\widebar{F}_{\tilde{q}_{d_i,j}}$:
\vspace{-7pt}
\begin{equation}\small
	\widetilde{\mathcal{J}}_{\tilde{q}_{d_i,j},\widebar{F}_{\tilde{q}_{d_i,j}}} = \bigcup_{c=1}^{m} LLM(P_{judg}(\tilde{q}_{d_i,j}, \tilde{d}_{c})).
	\vspace{-7pt}
\end{equation}
\noindent where $\tilde{d}_c$ is a document in the fused ranking $\widebar{F}_{\tilde{q}_{d_i,j}}$. We then can collate\footnote{Via the set union operation.} pseudo-relevance judgements across all queries for a sample document, and all sample documents for the target corpus $D_\mathcal{T}$, obtaining the relevance judgements set $\widetilde{\mathcal{J}}_\mathcal{T}$ (or $\widetilde{\mathcal{J}}$ for brevity of notation, since we consider only one target collection $\mathcal{T}$).

 To ``implement'' $P_{judg}$ we adapt the fine-grained relevance labels generation prompt of~\citet{zhuang2023beyond} because of its previously reported high effectiveness\footnote{We note other prompts for query-document relevance judgements have been proposed, e.g., that of \citet{thomas2023large}; we adapted the one of \citet{zhuang2023beyond} over others because of its simplicity. We leave evaluating the impact on LARMOR of alternative prompts for relevance evaluation to future work.}. As an example, our relevance judgment prompt for the NQ collection is \textit{``For the following query and document, judge whether they are `Highly Relevant', `Somewhat Relevant', or `Not Relevant'. Query: $\{\tilde{q}_j\}$ Document: $\{\tilde{d_{c}}\}$"}. Following \citet{zhuang2023beyond}, we only consider $\tilde{d_{c}}$ to be relevant to $\tilde{q}_j$ if the LLM generates \textit{`Highly Relevant'}, i.e., we convert the graded judgement to binary judgments. Finally $\widetilde{\mathcal{J}}$ 
  will be used for ranking DRs, which we discuss in details in the next section.
 
Prompt $P_{judg}$ guides the LLM to generate relevance judgments at a document level. In addition, we propose a second prompt, $P_{rank}$, to evaluate DRs at the ranking level. $P_{rank}$ is designed to instruct the LLM to generate a highly effective document ranking $\widebar{L}_{\tilde{q}_{d_i,j}}$ to be used as pseudo-reference list for the generated query $\tilde{q}_{d_i,j}$.
A reference list is commonly used in QPP~\cite{Shtok2016QueryPP}, giving rise to effective predictive methods in that context. In our work, we adapt this idea to enhance our approach.
This is achieved by prompting the LLM to re-rank the $m$ documents in the fused ranking for query $\tilde{q}_{d_i,j}$, i.e., $\widebar{F}_{\tilde{q}_{d_i,j}}$:
\vspace{-10pt}
\begin{equation}\small
\widebar{L}_{\tilde{q}_{d_i,j}} = LLM(P_{rank}(\widebar{F}_{\tilde{q}_{d_i,j}})
\vspace{-4pt}
\end{equation}


To ``implement'' $P_{rank}$ we used the Setwise document ranking prompt~\cite{zhuang2023setwise}. We note other LLM ranking prompts could have been used, e.g., Pointwise~\cite{zhuang-etal-2023-open, sachan-etal-2022-improving, zhuang2023beyond}, Listwise~\cite{sun-etal-2023-chatgpt, ma2023zero, pradeep2023rankvicuna, pradeep2023rankzephyr, tamber2023scaling}, Pairwise~\cite{qin2023large}. We chose the Setwise prompt because of its robustness and high effectiveness; due to space and computation constraints, we leave the study of other prompts for implementing $P_{rank}$ and their impact on LARMOR to future work.

Finally, we collate all reference lists into a set of reference list $\widetilde{\mathcal{L}}_\mathcal{T}$ (or for simplicity of notation, $\widetilde{\mathcal{L}}$) for the target corpus $\mathcal{T}$, keeping track of which generated query each list refers to.

\vspace{-4pt}
\subsection{Dense Retriever Ranking}
The final step in our LARMOR pipeline is ranking all the dense retrievers in the candidate pool using either or both of the generated pseudo-relevance judgments $\widetilde{\mathcal{J}}$ and the pseudo-reference lists $\widetilde{\mathcal{L}}$.

To rank DRs for a target collection with the pseudo-relevance judgements, we first produce document rankings in answer to the generated queries using all the DRs, and we then evaluate each of these rankings using the target evaluation measure $\mathcal{E}$ (in our empirical evaluation, $\mathcal{E} =$ nDCG@10). We then average the evaluation values across all queries to associate an estimated average evaluation measure $\widetilde{\mathcal{E}}$ to each of the dense retrievers (it is estimated because pseudo queries and judgements are used, in place of the real ones from the target collection). Subsequently we rank DRs in descending order of $\widetilde{\mathcal{E}}$.  

To rank DRs for a target collection with the reference lists $\widetilde{\mathcal{L}}$, we calculate the average Rank Bias Overlap (RBO)~\cite{webber2010rbo} of the rankings obtained using each DR for each generated query $\tilde{q}_{d_i,j}$ against its corresponding pseudo-reference list $\widebar{L}_{\tilde{q}_{d_i,j}}$. We then rank DRs in descending order of RBO values.

Finally, as the above two rankings of dense retrievers have been obtained leveraging different relevance signals (judgements vs. reference lists), we posit it is beneficial to combine these rankings to obtain a comprehensive and effective ordering of the DRs, from the one thought to be most effective on the target corpus to the least effective. Thus, we further employ a fusion algorithm to merge the rankings of dense retrievers\footnote{We further stress that in this step we fuse rankings of DRs, and not of documents like when creating reference lists in Section~\ref{sec:psuedo-jud-list}.} and create our final ranking of DRs used as solution to the dense retrievers selection task.

 \section{Experimental Setup}
\label{sec:setup}
We aim to comprehensively evaluate our proposed DR selection approach against a large pool of state-of-the-art DRs across a wide range of corpora from different domains. In this section, we outline the details of the criteria for selecting DRs from the MTEB leaderboard, along with the corpora used in our experiments. Finally, we provide the implementation and evaluation details of our approach.

\vspace{-6pt}
\subsection{Dense Retriever Pool and Target Corpora}

We assembled a large collection of state-of-the-art dense retrievers through the following steps:

We began by examining the MTEB leaderboard, selecting the top 30 retrievers based on their average performance across all corpora featured in the MTEB benchmark. Next, we assessed the performance of the retrievers on each individual corpus, expanding our set to include any retrievers that ranked in the top 30 for a specific corpus but were not part of our initial overall selection. This approach naturally led to overlapping models, as those most effective on one corpus often performed well on others. However, certain models demonstrated unique strengths in specific corpora. For instance, the \texttt{all-mpnet-base-v2} model is ranked the best for both SciDocs and FiQa corpora, yet it is 48th overall. Following the selection of leading models for each corpus, we refined our pool to align with our budgetary and computational constraints. This entailed removing API-based retrievers, e.g. Cohere, and any models with more than 6B parameters.

This process ultimately resulted in a carefully curated pool of 47 state-of-the-art dense retrieval models, which we will be using for all the experiments throughout the paper. Note that the number of models in our study substantially exceeds those utilized by Khramtsova et al.~\cite{khramtsova2023}, thereby increasing the complexity of the task and enhancing the credibility of our results.


For evaluation corpora, we follow Khramtsova et al.~\cite{khramtsova2023} who utilized the corpora from the BEIR benchmark~\cite{thakur2021}, which is widely employed for zero-shot dense retriever evaluation. This benchmark includes 18 collections across 9 diverse tasks. In accordance with standard practice, we selected a representative subset of 13 corpora, covering all 9 tasks featured in BEIR. The primary advantage of employing this benchmark is that none of its corpora were explicitly used for training the dense retrievers in our model pool. This makes it an appropriate choice for an unsupervised model selection task, especially in scenarios with domain shift between training and test.

\vspace{-6pt}
\subsection{Implementation Details}

We employ LLMs in our proposed DR selection pipeline to generate queries, pseudo-relevance judgments, and pseudo-reference lists. We consistently use the FLAN-T5~\cite{chung2022scaling} LLM through out the pipeline since it demonstrated strong effectiveness in zero-shot query generation~\cite{dai2022promptagator} and document ranking~\cite{zhuang2023setwise,qin2023large,zhuang-etal-2023-open}.

For the query generation component, inspired by recent works of task-aware dense retriever training~\cite{asai-etal-2023-task,wang2022text,zhang2023retrieve}, we adapted their prompts to the task of query generation to generate in-domain pseudo-relevant queries for each BEIR corpus. Specifically, our query generation prompt templates have two key pieces of knowledge related to the target domain. Firstly, we specify the type of target query to generate, such as questions for question answering or arguments for argument retrieval. Secondly, we identify the type of document, distinguishing between sources like Wikipedia pages and scientific titles with abstracts. Due to space constraints, we direct readers to refer to our github repository for the list of our query generation prompts~\footnote{https://github.com/ielab/larmor/blob/main/prompts.py} and the resulting generated queries for each corpus ~\footnote{https://github.com/ielab/larmor/tree/main/generated\_data/}.
We randomly sample $k=100$ documents from the target corpus and employ top-p sampling with $p=0.9$ to generate 10 queries for each sampled document, resulting in $|\widetilde{Q}_{\mathcal{T}}| = 1000$ generated queries per corpus. In Section~\ref{sec:num_q}, we investigate the impact of the number of generated queries per document as well as the influence of using different backbone LLMs in query generation.

Regarding the prompt for generating pseudo-relevance judgments, we modify the fine-grained relevance label generation prompts \cite{zhuang2023beyond} to align with our domain-specific query generation prompts. This involves incorporating information about the query type and document type into the prompts. We then use the prompt to instruct LLMs to judge the top $m=100$ documents from the $\widebar{F}_{\tilde{q}_{d_i,j}}$ ranking for each generated query. We again refer readers to our github repository for the details of our prompts. As for the generation of pseudo-reference lists $\widebar{L}_{\tilde{q}_{d_i,j}}$, we simply employ the original Setwise ranking prompt proposed by \citeauthor{zhuang2023setwise}~\cite{zhuang2023setwise} with the default setting of using heap sort algorithm and compare 3 documents at a time to re-rank the top $m=100$ documents from the $\widebar{F}_{\tilde{q}_{d_i,j}}$. 

For the fusion algorithm used in LARMOR to create $\widebar{F}_{\tilde{q}_{d_i,j}}$ and the final DR ranking, considering that the scores provided by different DRs might have different scales, we opt for Reciprocal Rank Fusion (RRF)\cite{gordon2009rrf}, a position-based method. We employ the implementation from the Python toolkit ranx\cite{ranx} with the default parameters.

\begin{table*}[t]
	\caption{Kendall Tau Correlation value, calculated based on nDCG@10. \vspace{-10pt}}
	\resizebox{\textwidth}{!}{
		\centering
		\begin{tabular}{c|c|c|c|c|c|c|c|c|c|c|c|c|c|c}
			\hline
			&  NF & FiQA & ArguAna & SciDocs & SciFact& Covid & Quora&  NQ & DBPedia & HotpotQA  & Signal1M & Robust04 & Trec-News &\textbf{Avrg}\\
			\hline
			\texttt{MSMARCO perf}. &0.337 & 0.240 & 0.17 & 0.118 & 0.291 & 0.339 & 0.298 & \textbf{0.646} & 0.515 & 0.492 & 0.121 & 0.141 & 0.186 & 0.300\\
			\texttt{Binary Entropy} &-0.056 & -0.164 & -0.048 & -0.086 & 0.103 & -0.183 & -0.280 & -0.081 & 0.034 & 0.165 & 0.069 & -0.106 & -0.212 & -0.065 \\ 
			\texttt{Query Alteration} & -0.277 & -0.250 & -0.173 & -0.242 & -0.199 & -0.152 & -0.195 & -0.458 & -0.359 & -0.217 & 0.029 & -0.102 & -0.194 & -0.215\\
			
			\hline
			
			\texttt{WIG} & -0.156 & -0.212 & -0.093 & -0.212 & -0.188 & -0.231 & -0.201 & -0.398 & -0.262 & -0.149 & 0.112 & -0.103 & -0.191 & -0.176 \\
			\texttt{WIG Norm} & -0.027 & -0.125 & -0.138 & -0.105 & -0.001 & -0.118 & -0.147 & 0.078 & 0.053 & -0.049 & 0.010 & -0.106 & -0.055 & -0.056 \\
			\texttt{NQC} & -0.036 & -0.021 & 0.142 & 0.08 & -0.01 & -0.202 & -0.071 & -0.304 & -0.153 & -0.086 & 0.036 & 0.051 & -0.006 & -0.045 \\
			\texttt{NQC} Norm & -0.136 & -0.191 & -0.06 & -0.121 & -0.136 & -0.198 & -0.197 & -0.441 & -0.262 & -0.154 & 0.099 & -0.080 & -0.160 & -0.157 \\
			\texttt{SMV} &-0.056 & -0.012 & 0.143 & 0.056 & 0.010 & -0.198 & -0.066 & -0.289 & -0.162 & -0.110 & 0.036 & 0.047 & 0.005 & -0.046 \\
			\texttt{SMV Norm} & -0.173 & -0.179 & -0.066 & -0.136 & -0.127 & -0.204 & -0.191 & -0.429 & -0.280 & -0.197 & 0.103 & -0.075 & -0.154 & -0.162 \\
			$\sigma$ & -0.204 & -0.228 & -0.116 & -0.149 & -0.186 & -0.198 & -0.142 & -0.402 & -0.260 & -0.219 & 0.084 & -0.099 & -0.167 & -0.176 \\
			$\sigma_{max}$& -0.147 & -0.236 & -0.123 & -0.114 & -0.182 & -0.224 & -0.236 & -0.370 & -0.291 & -0.166 & 0.062 & -0.064 & -0.227 & -0.178 \\
			\texttt{Clarity} & 0.114 & 0.245 & 0.223 & 0.038 & 0.264 & -0.333 & 0.345 & 0.059 & -0.186 & -0.145 & 0.203 & 0.08 & -0.193 & 0.055\\
			
			\texttt{Fusion} & 0.653 & 0.436 & 0.544 & 0.686 & 0.636 & 0.368 & 0.670 & 0.374 & \textbf{0.719 }& 0.555 & \textbf{0.506} & \textbf{0.611 }& \textbf{0.698} & 0.574\\          
			\hline
			
			LARMOR (ours) & \textbf{0.700} & \textbf{0.618} & \textbf{0.627} & \textbf{0.739} & \textbf{0.766} & \textbf{0.553} & \textbf{0.740} & 0.563 & 0.665 & \textbf{0.710} & 0.380 & 0.444 & 0.690 & \textbf{0.630} \\
        
			\hline
			
		\end{tabular}
	}
	\label{tab:tau_ndcg@10}
	\vspace{-4pt}
\end{table*}

\vspace{-6pt}

\subsection{Evaluation}
For evaluating our proposed LARMOR and baselines, we follow previous work that uses Kendall Tau correlation and $\Delta_e$ to assess the performance of methods on the DR selection task.

Both evaluations require the DR's ground truth performance ranking on the target corpus. Therefore, for each of the considered corpora, we rank the DRs based on the nDCG@10 obtained from the test queries and human judgments provided by each collection. This score is the official evaluation measure for BEIR.

After obtaining the ground truth DR performance ranking, Kendall Tau correlation is used to assess the similarity between the rankings generated by the DR selection methods and the ground truth ranking. Specifically, Kendall Tau correlation measures the proportion of document pairs that are ranked in the same order by both rankings. A score of 1 indicates perfect positive correlation, -1 indicates perfect negative correlation, and 0 indicates completely random correlation.

On the other hand, $\Delta_e$ aims to measure the performance gap between the selected DR and the ground truth best-performing DR for a specific DR evaluation measure $e$. This is defined as:
\vspace{-4pt}
\begin{equation}
	\Delta_e = e(M(\mathcal{R})) - e(R^*)
	\vspace{-4pt}
\end{equation}
where $R^*$ is the ground truth best DR, and $M(\mathcal{R})$ is the DR ranked at top by method $M$. In our experiments, we set $e$ to be nDCG@10 to align with the target DR performance measurement. If $\Delta_e = 0$, it means that method $M$ successfully ranked the best DR at the top.

\vspace{-6pt}
\subsection{Other Baselines}

We briefly describe the baselines used for comparison.

\noindent\textbf{Model Selection Methods~\cite{khramtsova2023}}:

\begin{itemize}
    \item \textit{MSMARCO performance}: ranks models based on their performance on MSMARCO - a large widely-used public collection for fine-tuning and evaluating retrieval models.
    \item \textit{Binary Entropy} evaluates model uncertainty. For each query, the entropy of the probability-at-rank distribution is calculated. DRs are then ranked based on the average entropy across all queries.
    \item \textit{Query Alteration} assesses the sensitivity of DRs to query variations. It involves modifying the query and measuring the standard deviation of scores for the retrieved documents. DRs are ranked based on the average standard deviation across queries, with smaller values indicating greater robustness against query perturbation, thereby implying higher retrieval credibility.
\end{itemize}

\noindent\textbf{QPP-based methods:}

\begin{itemize}
    \item Weighted Information Gain (WIG)~\cite{Yun2007Wig} measures the disparity between the average retrieval scores of top-ranked documents and the overall score of the corpus.  
    \item Normalized Query Commitment (NQC)~\cite{Shtok2012NQC} calculates the standard deviation of the scores of top-ranked documents.
    \item Scores Magnitude and Variance (SMV)~\cite{Tao2014SMV} combines both the magnitude and the standard deviation of the scores of top-ranked documents.

    \item Clarity~\cite{Cronen2002Clarity} measures the discrepancy between the language model built from the top retrieved results and the language model of the entire corpus.
    
    \item $\sigma$ ~\cite{Perez2010Sigma} calculates the standard deviation of scores, determining the optimal number of retrieved documents for each query to minimize the impact of low-scoring, non-relevant documents. 
    \item $\sigma_{\max}$~\cite{Cummins2011SigmaMax} is a normalized standard deviation that considers only documents with scores above a certain percentage of the top score.

    \item Fusion~\cite{Shtok2016QueryPP} relies on submitting the target query to all candidate DRs to acquire document rankings for each DR. A search result fusion method is then used to aggregate these rankings into a pseudo-reference list. DRs are subsequently scored based on their RBO against this reference list.
\end{itemize}

For all QPP-based methods, the final score of a DR with respect to a target collection is computed as the average value returned by the QPP method across all queries.

\vspace{-6pt}
\section{Results}
\vspace{-2pt}

\begin{table*}[t]
    \caption{$\Delta_e$, calculated based on nDCG@10. \vspace{-10pt}}
\resizebox{\textwidth}{!}{
    \centering
    \begin{tabular}{c|c|c|c|c|c|c|c|c|c|c|c|c|c|c}

        \hline
        &  NF & FiQA & ArguAna & SciDocs & SciFact& Covid & Quora&  NQ & DBPedia & HotpotQA  & Signal1M & Robust04 & Trec-News &\textbf{Avrg}\\
        \hline
        \texttt{MSMARCO Perf.} & 6.84 & 20.21 & 27.47 & 10.02 & 20.26 & 32.18 & 9.58 & 13.91 & 9.80 & 17.52 & 3.51 & 20.23 & 16.89 & 16.03\\
        \texttt{Binary Entropy} & 6.84 & 18.54 & 25.84 & 11.59 & 25.06 & 9.10 & 2.21 & 34.34 & 6.26 & 29.2 & 5.48 & 7.59 & 12.31 & 14.95 \\
        \texttt{Query Alteration} & 2.61 & 15.15 & 7.98 & 6.04 & 18.35 & 13.42 & 3.61 & 21.04 & 2.46 & 20.15 & 4.50 & 7.59 & 5.95 & 9.91\\         
        \hline
        \texttt{WIG} & 7.06 & 16.69 & 25.92 & 6.60 & 8.44 & 12.16 & 1.07 & 18.18 & 4.56 & 20.15 & 5.69 & 15.85 & 13.06 & 11.96 \\        
        \texttt{WIG Norm} &  13.46 & 24.25 & 24.45 & 12.42 & 22.39 & 14.58 & 3.98 & 17.36 & 13.38 & 29.74 & 1.82 & 9.22 & 14.82 & 15.53 \\
        
        \texttt{NQC} & 15.09 & 20.48 & 23.22 & 11.59 & 25.05 & 19.59 & 4.04 & 19.68 & 16.79 & 29.2 & 4.14 & 15.28 & 12.31 &16.65 \\ 
        \texttt{NQC Norm} & 7.06 & 16.69 & 25.92 & 6.60 & 8.44 & 15.17 & 4.29 & 7.77 & 13.34 & 20.15 & 5.69 & 15.85 & 13.06 & 12.32 \\
        
        \texttt{SMV} & 15.09 & 20.48 & 23.22 & 11.59 & 25.05 & 19.59 & 4.04 & 19.68 & 16.79 & 29.2 & 4.14 & 15.28 & 12.31 & 16.65 \\
        \texttt{SMV Norm} & 7.61 & 15.15 & 25.92 & 9.83 & 8.44 & 15.17 & 4.29 & 7.77 & 13.34 & 7.73 & 5.69 & 15.85 & 13.06 & 11.53 \\
        
        $\sigma$ & 7.61 & 16.69 & 25.92 & 6.60 & 10.58 & 15.17 & 4.29 & 4.96 & 3.60 & 20.15 & 4.50 & 15.85 & 10.36 & 11.25 \\
        $\sigma_{max}$ & 7.61 & 16.69 & 11.45 & 6.67 & 5.74 & 15.17 & 2.20 & 21.04 & 6.26 & 18.64 & 3.62 & 15.85 & 10.36 & 10.87 \\
        \texttt{Clarity} & 2.44 & 12.46 & 18.59 & 2.12 & 11.62 & 33.56 & 3.52 & 4.06 & 12.8 & 35.04 & 2.00 & 23.43 & 7.84 & 13.04 \\
        
        \texttt{Fusion} &\textbf{1.15} & 5.46 & 7.96 & 5.09 & 3.67 & 20.09 & \textbf{0.94} & 9.29 & 2.53 & \textbf{7.17} & \textbf{0.53 }& \textbf{2.37} & \textbf{0.0} & 5.10 \\          
        
        \hline

        LARMOR (ours) & 2.44 & \textbf{3.07} & \textbf{0.0 }& \textbf{0.79 }& \textbf{0.77 }& \textbf{6.81} & \textbf{0.94} & \textbf{0.58} & \textbf{0.82 }& \textbf{7.17} & 1.28 & \textbf{2.37} & \textbf{0.0} & \textbf{2.08} \\

        \hline
        
        \hline
    \end{tabular}
}
\vspace{-6pt}
\label{tab:delta}
\end{table*}

\begin{table*}[t]
	\caption{Ablation Study: the effect of different steps of the pipeline. Kendall Tau Correlation value, calculated based on nDCG@10. \vspace{-10pt}}
\resizebox{0.9\textwidth}{!}{
	\centering
	\begin{tabular}{c|c|c|c|c|c|c|c|c|c|c|c|c|c|c}
		\hline
		&  NF & FiQA & ArguAna & SciDocs & SciFact& Covid & Quora&  NQ & DBPedia & HotpotQA  & Signal1M & Robust04 & Trec-News &\textbf{Avrg}\\
		\hline
		Q & 0.483 & 0.525 & 0.522 & 0.545 & 0.708 & 0.563 & 0.635 & 0.578 & 0.658 & 0.824 & 0.153 & 0.324 & 0.634 & 0.550 \\
		QF & 0.677 & 0.545 & 0.502 & 0.729 & 0.610 & 0.313 & 0.622 & 0.311 & 0.648 & 0.520 & 0.517 & 0.539 & 0.648 & 0.552\\
		QFJ &  0.552 & 0.562 & 0.522 & 0.59 & 0.809 & 0.559 & 0.677 & 0.646 & 0.667 & 0.789 & 0.191 & 0.397 & 0.648 & 0.585\\
		QFR & 0.700 & 0.562 & 0.448 & 0.764 & 0.676 & 0.370 & 0.672 & 0.444 & 0.613 & 0.607 & 0.526 & 0.437 & 0.667 & 0.576\\
		LARMOR & 0.700 & 0.618 & 0.627 & 0.739 & 0.766 & 0.553 & 0.740 & 0.563 & 0.665 & 0.710 & 0.380 & 0.444 & 0.690 & 0.630 \\

		\hline
		
	\end{tabular}
}
\vspace{-4pt}
\label{tab:abl_tau_ndcg}

\end{table*}

\begin{table*}[t]
	\caption{Ablation Study: the effect of different steps of the pipeline. $\Delta_e$, calculated based on nDCG@10. \vspace{-10pt}}
\resizebox{0.9\textwidth}{!}{
	\centering
	\begin{tabular}{c|c|c|c|c|c|c|c|c|c|c|c|c|c|c}
		\hline
		&  NF & FiQA & ArguAna & SciDocs & SciFact& Covid & Quora&  NQ & DBPedia & HotpotQA  & Signal1M & Robust04 & Trec-News &\textbf{Avrg}\\
		\hline
		
		Q & 1.52 & 5.46 & 13.73 & 3.26 & 2.01 & 18.29 & 0.0 & 0.58 & 3.87 & 3.10 & 1.38 & 7.92 & 5.95 & 5.16 \\
		QF &0.22 & 5.46 & 7.96 & 0.79 & 3.67 & 18.31 & 0.94 & 9.29 & 2.53 & 7.17 & 0.25 & 2.37 & 0.0 & 4.53 \\          
		QFJ & 1.74 & 3.07 & 1.35 & 3.90 & 0.77 & 5.31 & 0.34 & 0.58 & 4.56 & 3.10 & 1.38 & 7.92 & 5.95 & 3.07 \\ 
		QFR & 0.22 & 5.46 & 8.08 & 0.79 & 0.77 & 14.67 & 0.94 & 8.21 & 3.73 & 7.17 & 0.25 & 2.37 & 0.0 & 4.05\\ 
		
		LARMOR &2.44 & 3.07 & 0.0 & 0.79 & 0.77 & 6.81 & 0.94 & 0.58 & 0.82 & 7.17 & 1.28 & 2.37 & 0.0 & 2.08\\

		\hline
		
	\end{tabular}
}
\vspace{-7pt}
\label{tab:abl_delts}

\end{table*}

\subsection{Main Results}

In Tables~\ref{tab:tau_ndcg@10} and~\ref{tab:delta}, we compare LARMOR against other baselines in terms of Kendall Tau and $\Delta_e$, respectively.

Firstly, we observe that the baseline \texttt{MSMARCO pref}, which simply ranks DRs based on nDCG@10 obtained on the MSMARCO collection, performs the best among the previous DR selection methods in terms of Kendall Tau; this finding aligns with previous work~\cite{khramtsova2023}. However, it is important to note that it is not guaranteed that MS MARCO training data is used in all the DRs we considered in our experiments.
For example, \texttt{bge-large-en-v1.5} was trained on the Massive Text Pairs (MTP) collection, which contains 200M English text pairs~\cite{bge_embedding}. It is the second-best performing DR across all collections, however it is predicted to be only the 8-th best if one relies on \texttt{MSMARCO pref}, being surpassed by DRs that likely overfit the MSMARCO collection but do not perform comparably well on the other collections.
It is noteworthy that the prediction based on \texttt{MSMARCO pref} yields the best Kendall Tau for the NQ collection. This is somewhat expected since NQ is considered to be the most similar collection to MSMARCO.

Another performance-based approach, \texttt{Query Alteration}, closely follows \texttt{MSMARCO pref}. While it underperforms compared to \texttt{MSMARCO pref} in terms of Kendall Tau, it achieves a higher average $\Delta_e$, indicating its greater effectiveness in selecting the top DRs rather than  providing the true ranking of all DRs.

As expected, methods that rely on comparing the scores produced by the retrievers perform poorly in both DR selection and DR ranking tasks. These methods include \texttt{Binary Entropy} and four QPP methods (WIG, NQC, SMV, $\sigma$). 
Notably, the normalized versions of score-based QPPs yield better results in terms of both $\Delta_e$ and Kendall Tau (Tables \ref{tab:tau_ndcg@10} and \ref{tab:delta}).  This implies that incorporating a scaling parameter for normalizing scores is beneficial.  However, as discussed in Section ~\ref{sec:baseline_challenges}, this normalization primarily regularizes scores within the collection, but does not address the challenge of score distribution diversity across different DRs.

In contrast, \texttt{Fusion}, which aggregates the retrieved lists from different DRs without relying on absolute score values, achieves significantly better performance in terms of both Kendall Tau and $\Delta_e$. Nevertheless, similar to the other QPP baselines, \texttt{Fusion} requires the availability of the queries from the target collection. This requirement is often impractical, as DR selection must occur before the system is deployed and queries are gathered. 
It is important to note that our LARMOR, unlike other baselines, is query-free.

Finally, our LARMOR achieves the best overall performance in terms of both Kendall Tau and $\Delta_e$. Moreover, LARMOR selects the best DR for two collections (Arguana and Trec-News),  resulting in an average nDCG drop of only 2.08\% across the board.

In the next subsections we tease out the contributions of LARMOR's components, and the effect of the LLM model size, of the specific LLM backbone, and of the number of generated queries. 

\vspace{-8pt}
\subsection{Effect of LARMOR's Components}
\label{sec:ablation}
We study the effectiveness of some of LARMOR's components, measuring effectiveness at intermediate ablation points in the pipeline, illustrated in Figure~\ref{pipeline}. The ablation points we investigate are:

\begin{itemize}
	\item \textbf{Q:} Query generation --- we use the LLM-generated query, along with its associated document, as a pseudo-relevant query-document pair. The DRs are then evaluated and ranked based on these judgments. Note that here we only have a single relevant document for each generated query.
	
	\item \textbf{QF:} Rank fusion of the generated queries --- we generate multiple queries for each sampled document $d_i$, and fuse their ranking to obtain $\widebar{F}_{\tilde{q}_{d_i,j}}$ for each $d_i$. We then rank DRs based on the average RBO value against the obtained set of fused rankings $\widebar{F}_{\tilde{q}_{d_i,j}}$.
	
	\item \textbf{QFJ:} Pseudo-judgments from rank fusion --- we generate multiple queries for each sampled document $d_i$, and fuse their ranking to obtain $\widebar{F}_{\tilde{q}_{d_i,j}}$ for each $d_i$. We then use the LLM to generate pseudo-judgements ($\widetilde{\mathcal{J}}$), and use these to rank DRs.
	
	\item \textbf{QFR:} Pseudo-reference lists from rank fusion --- using the same $\widebar{F}_{\tilde{q}_{d_i,j}}$ described above, we use the LLM to re-rank $\widebar{F}_{\tilde{q}_{d_i,j}}$, obtaining the pseudo-reference list $\widebar{L}_{\tilde{q}_{d_i,j}}$. We then rank DRs with respect to the set of all reference lists $\mathcal{L}$.
	
\end{itemize}
The full LARMOR pipeline performs an additional fusion of the ranking of DRs obtained from the ablation points QFJ and QFR. 

In Table~\ref{tab:abl_tau_ndcg} and~\ref{tab:abl_delts}, we present Kendall Tau and $\Delta_e$ result, obtained at these different steps of our pipeline.

As the results illustrate, with just judgments from Q, we can already achieve very strong performance that, outperforming most baselines, only falling short of QPP fusion.

QF further improves Q by using the fused ranking and RBO to rank DRs. We note that QF is similar to the QPP \texttt{Fusion} baseline except that the queries for QF are LLM-generated, while the queries for \texttt{Fusion} are the actual test queries. 
Remarkably, QF can surpass QPP \texttt{Fusion} in terms of $\Delta_e$ average score, suggesting that our LLM-generated in-domain queries are of satisfactory quality.

On the other hand, QFJ and QFR prove to be very important; they both significantly improve QF for both Kendall Tau and $\Delta_e$.

Finally, our whole pipeline LARMOR achieved the overall best performance by fusing QFJ and QFR. 
These results demonstrate that each component in LARMOR has a significant contribution.

\vspace{-6pt}
\subsection{Effect of LLM Model Size}

\begin{figure}
	\centering
	\includegraphics[width=0.235\textwidth]{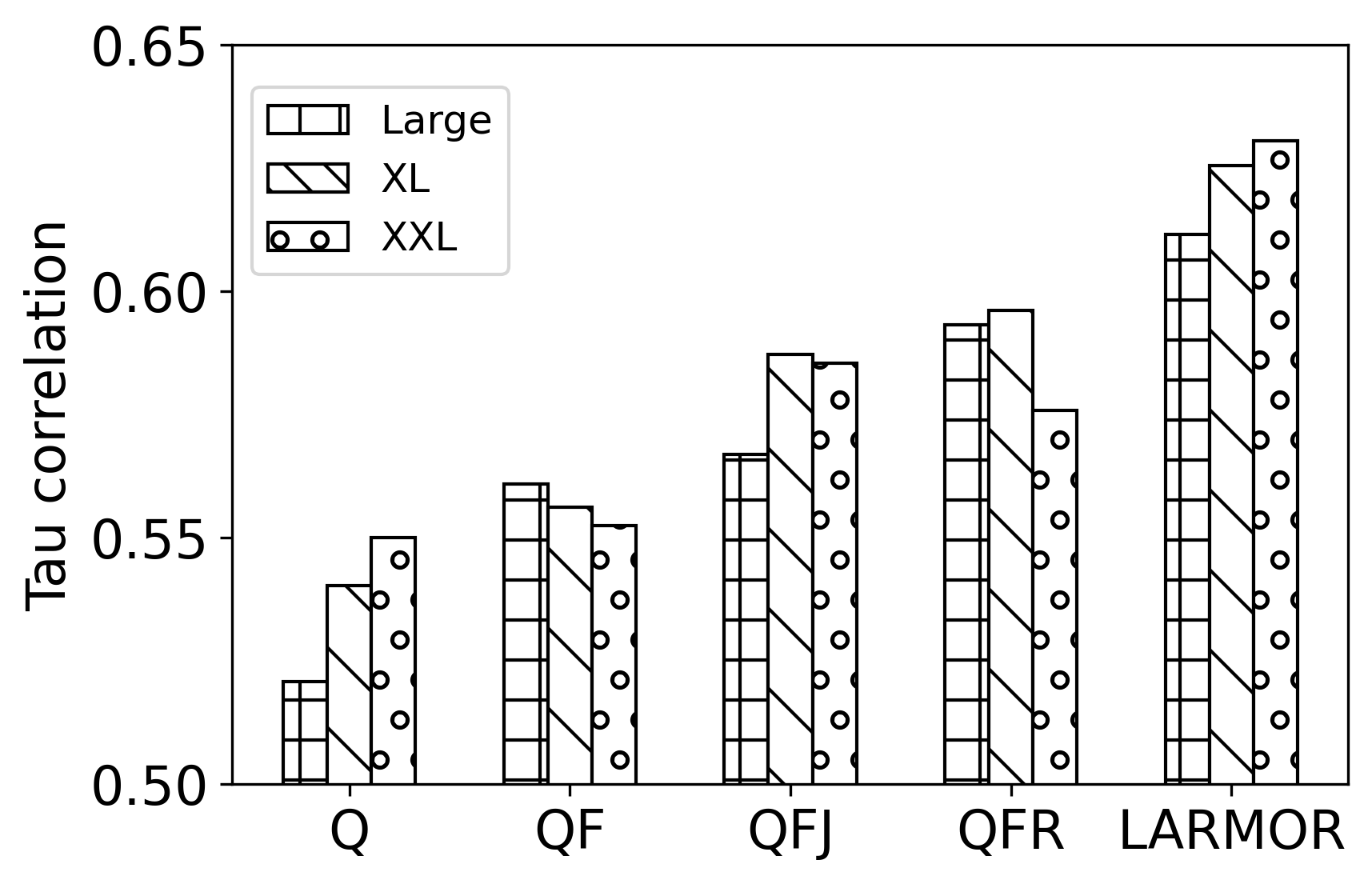}
	\includegraphics[width=0.22\textwidth]{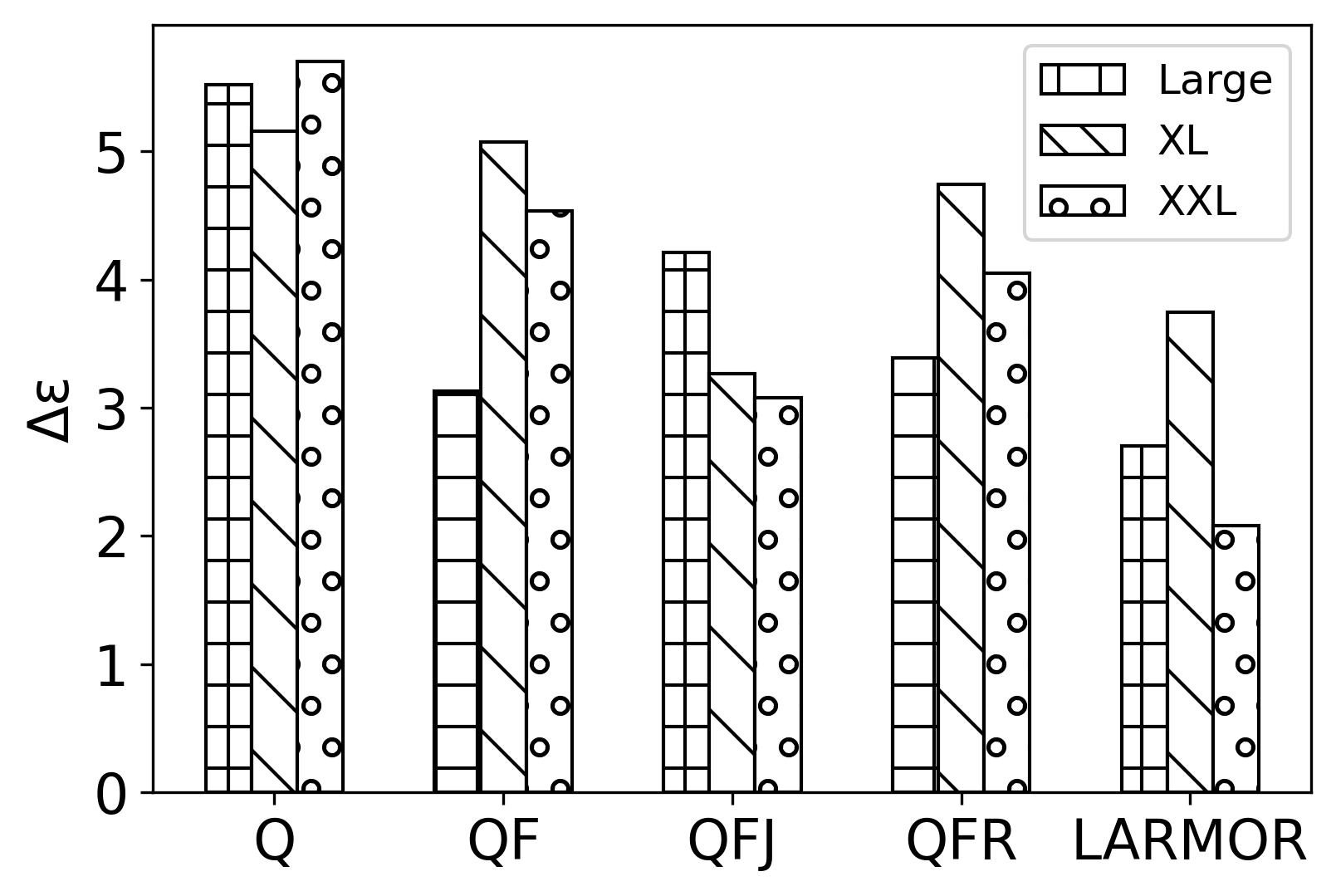}
	 \vspace{-4pt}
	\caption{Kendall Tau (left) and $\Delta_e$ (right) performance of our proposed LARMOR and its various components using different sizes of FlanT5 models.  \vspace{-6pt}
	}
	\label{barplot_steps}
	 \vspace{-10pt}
\end{figure}

To comprehensively understand the impact of LLM size on our proposed LARMOR, in Figure~\ref{barplot_steps} we plot the Kendall Tau and $\Delta_e$ performance across different steps of the pipeline. We explore the influence of FlanT5 models with varying sizes, namely FLAN-T5-large (780M), FLAN-T5-XL (3B), and FLAN-T5-XXL (11B).

Firstly, a clear pattern emerges as each step within the pipeline consistently contributes to improved effectiveness, irrespective of the model size with the only exception that FLAN-T5-large exhibits a suboptimal $\Delta_e$ score on QF compared to Q.

On the other hand, the performance across different model sizes exhibits variations at each pipeline step. This  suggests that the conventional scaling law of LLMs might not apply here: it is not always the case that a larger model performs better for our method with FLAN-T5 models. Nevertheless, FLAN-T5-XXL achieved the best performance when the full LARMOR pipeline is applied.

\subsection{Effect of LLM Backbone and Number of Generated Queries}\label{sec:num_q}
\begin{figure}
	\centering
	\includegraphics[width=0.225\textwidth]{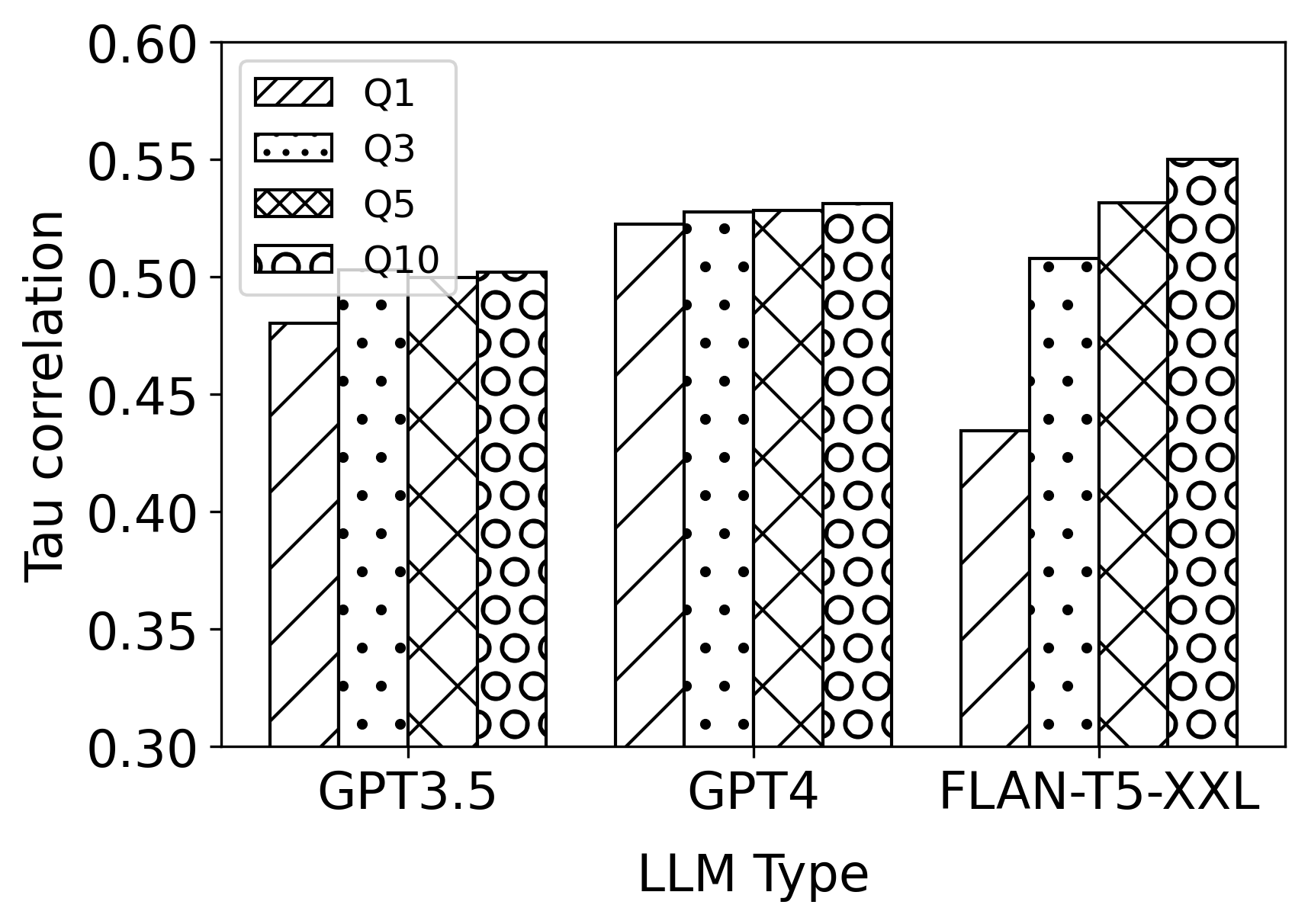}
    \includegraphics[width=0.21\textwidth]{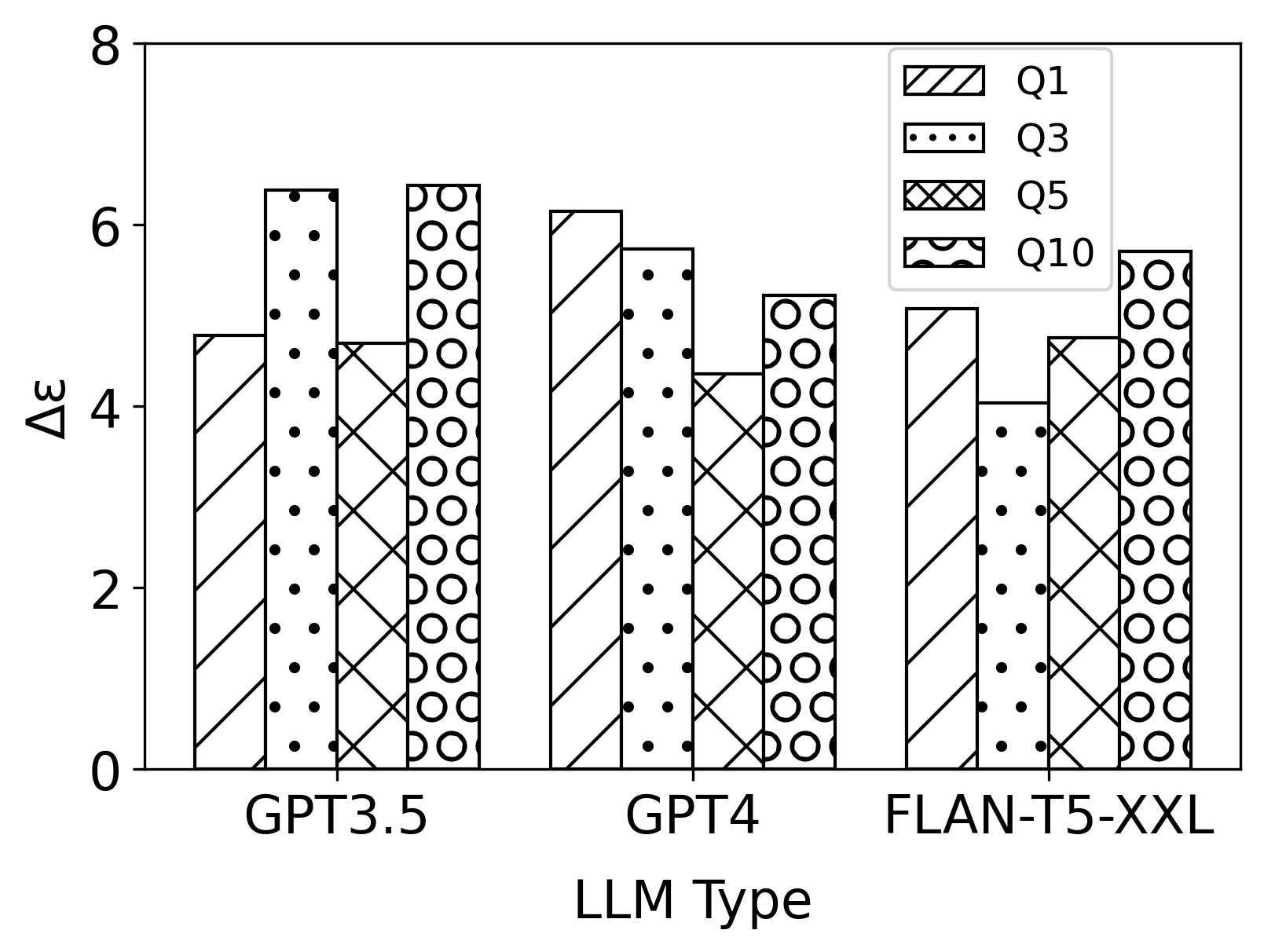}
    \vspace{-8pt}
	\caption{Kendall Tau (left) and $\Delta_e$ (right) performance of pseudo-query generation with a different number of generated queries and different LLM backbone.  \vspace{-6pt}
	}
	\label{barplot_size}
	\vspace{-10pt}
\end{figure}

In this section, we study how the state-of-the-art OpenAI LLMs, GPT-3.5 and GPT-4, perform compared to FLAN-T5-XXL. For these experiments, we only conduct tests on the query generation (Q) step due to the high cost of running the whole LARMOR with OpenAI models. Additionally, we investigate how the number of generated queries per document impacts performance. The results are illustrated in Figure~\ref{barplot_size}.

For Kendall Tau performance, it is evident that more generated queries per document tend to be beneficial, especially for FlanT5-XXL, although the improvements are marginal for GPT-3.5 and GPT-4. GPT-4 tends to have an overall high Kendall Tau score; however, FLAN-T5-XXL only outperforms when the number of generated queries is set to 10.
As for the $\Delta_e$ score, the impact of the number of generated queries varies, and the difference between models also varies.

However, we note that, although overall OpenAI models perform similarly to FLAN-T5-XXL, on the Arguana collection where FLAN-T5-XXL performs poorly (Kendall Tau = 0.522, $\Delta_e=13.73$), OpenAI models achieved surprisingly good performance. For example, GPT-4 achieved Kendall Tau = 0.713 and $\Delta_e=0$, which are the best scores on this collection. We observe that Arguana poses a non-trivial retrieval task—specifically, a counter-arguments retrieval task—requiring LLMs to have the capability to generate counter-arguments. GPT models demonstrate this capability effectively.

\section{Conclusion}

This paper introduces a novel LLM-based approach for dense retriever selection, the Large Language Model Assisted Retrieval Model Ranking (LARMOR). 
Dense retriever selection is a crucial task in many applications of search engines technologies. Dense retrievers are an effective and increasingly popular component of a search engine. 
Search engine practitioners are often faced with the choice of which dense retriever to deploy on a specific target collection. However, it is challenging to predict a DR's effectiveness on a target collection that contains data different from that in the collection used for training the DR. This is even more so if the practitioners do not have access to user queries and relevance judgements from the target collection, as it is often the case in many applications, e.g., in small-medium enterprises and in domains like health and legal, due to the cost and time required to collect these signals, or the impossibility to access this data for privacy reasons.

Notably, LARMOR stands out as the only available method that does not require any post-deployment data but instead relies on minimal prior knowledge about the target collection to design prompts to guide LLMs in generating synthetic queries, pseudo-relevant judgments, and reference lists. These in turn are used within LARMOR to rank dense retriever systems.

 We comprehensively evaluate LARMOR across 13 different BEIR collections, considering a large pool of state-of-the-art dense retrievers. Our results demonstrate that LARMOR accurately ranks DRs based on their effectiveness in a zero-shot manner, outperforming all previous DR selection methods and adapted QPP methods.

Notably, unlike many existing baselines (e.g., score-based QPP, \texttt{Query Alteration}, \texttt{Binary Entropy}), our method is model-agnostic and can be extended to choose among any type of IR models.

For future work, we are interested in applying advanced automatic prompt optimization methods~\cite{yang2023large,fernando2023promptbreeder} to further enhance the domain-specific prompts used by LARMOR. Additionally, we are also interested in incorporating into LARMOR and study recent advanced open-source LLMs, such as Mistral~\cite{jiang2023mistral} and Llama3~\cite{llama3modelcard}. 
Another promising avenue for future work is reducing the computational overhead of LLM-related computations in our pipeline.

\bibliographystyle{ACM-Reference-Format}
\balance
\bibliography{Bib/bibliography}

\end{document}